\documentclass[aps,twocolumn,superscriptaddress,showpacs,prl,amssymb, amsmath]{revtex4}

\newtheorem{defi}{Definition}
\newtheorem{lem}[defi]{Lemma}
\newtheorem{thm}[defi]{Theorem}
\def\QED{\mbox{\rule[0pt]{1.5ex}{1.5ex}}}

\def\endproof{\hspace*{\fill}~\QED\par\endtrivlist\unskip}

{\it}

\def\Tr{\mathop{\rm Tr}\nolimits}
\def\Id{\mathop{\rm I}\nolimits}

\def\SU{\mathop{\rm SU}}

\def\Label#1{\label{#1}\ [\ #1\ ]\ }
\def\Label{\label}

\begin{document}
\draft
\title{Universal distortion-free entanglement 
concentration}
\author{Masahito Hayashi}
\email{masahito@brain.riken.go.jp}
\affiliation{Laboratory for Mathematical Neuroscience, Brain Science Institute, RIKEN,
2-1 Hirosawa, Wako, Saitama, 351-0198, Japan}
\author{Keiji Matsumoto}
\email{keiji@qci.jst.go.jp}
\affiliation{Quantum Computation and Information Project, ERATO, JST,
5-28-3, Hongo, Bunkyo-ku, Tokyo, 113-0033, Japan}

\date{\today}
\begin{abstract}
Entanglement concentration from many copies of 
unknown pure states is discussed,
and we propose the protocol which 
not only achieves entropy rate, but also
produces the perfect maximally entangled state.
Our protocol is induced naturally 
from symmetry 
of $n$-tensored pure state,
and is optimal for all the protocols which concentrates 
entanglement from unknown pure states, in the sense of 
failure probability.
In the proof of optimality, the statistical estimation theory
plays a key role, for concentrated entanglement
gives a natural estimate of the entropy of entanglement.
\end{abstract}
\maketitle
Entanglement is a major source of wonder 
of quantum information processing.
In many quantum information processing,
for example teleporation, and dense coding, 
it is most desirable to share 
maximally entangled states ~\cite{B2,B1},
which makes worthwhile 
the study of {\it entanglement concentration}, or 
production  of a maximally entangled state 
from given partially entangled pure states through 
local operations and classical communications (LOCC). 

As proven by Bennett et al.~\cite{Ben}, 
when  $n\,(\gg 1)$ copies of $|\phi \rangle$ is 
shared by Alice and Bob, 
whose respective Hilbert spaces are denoted by
${\cal H}_A$ and ${\cal H}_B$ respectively, 
they can produce, through local operations, 
$2^{n H({\bf p}_\phi)}$-dimensional maximally entangled state 
with the probability $1$ asymptotically.
Here, 
${\bf p}_{\phi}= (p_{1,\phi}, \ldots, p_{d,\phi})$  are  
the Schmidt coefficients of $|\phi\rangle$, 
(i.e., 
$|\phi\rangle=\sum_i \sqrt{p_{i,\phi}}|e_{i,A}\rangle |e_{i,B}\rangle$), 
with $p_{1,\phi} \ge p_{2,\phi} \ge \ldots \ge p_{d,\phi}$,
$H({\bf p})$ is the Shannon entropy of ${\bf p}$,
and 
$k$-dimensional maximally entangle state means
the state such that 
$\frac{1}{\sqrt{k}}\sum_i|e_{A,i}\rangle|e_{B,i}\rangle$.
(Without loss of generality, 
${\cal H}_A={\cal H}_B=d$ is assumed.)

In this letter, we treat the case where $|\phi\rangle$ is unknown,
and the {\it perfect} (not approximate) entangled state is needed,
while the dimension of the obtained entangled state is probabilistic.
We propose a protocol $\{C_*^n\}$
which produces a  $2^{n H({\bf p}_\phi)}$-dimensional 
maximally entangled state asymptotically 
with the probability $1$ even in this difficult setting.
This kind of protocol is called 
a {\it universal distortion-free entanglement 
concentration},
while a protocol outputting {\it approximate}
$2^{n H({\bf p}_{\phi})}$-dimensional maximally entangled state 
is called {\it universal approximate entanglement concentration},
whose example  is straightforwardly constructed as follows:
perform the entanglement concentration protocol of
\cite{Ben}
after the estimation of the Schmidt basis of $|\phi\rangle$ by measuring 
$m$ ($n\gg m\gg 1$) copies of $|\phi\rangle$.
In this way, however, the final state is 
not quite a maximally entangled state, 
because of 
the noises caused by the errors in the estimation.

The difficulty 
of construction of a universal distortion-free concentration
mainly comes from the lack of the knowledge about 
the Schmidt basis:
indeed, if the Schmidt basis is known,
the protocol of \cite{Ben} is successfully applied to 
produce a perfect maximally entangled state.
This difficulty is overcome by focusing on the symmetry of 
the $n$-tensored pure state $|\phi\rangle^{\otimes n}$.

The merit of our protocol is not only the distortion-free property, 
but also the optimality for all the universal approximate 
concentrations, in the sense of smallest failure probability.
In the proof, aside from  arguments based on symmetry,
the statistical estimation theory plays a key role,
for concentrated entanglement gives a natural estimate of
$H({\bf p}_{\phi})$.
Remarkably,
our protocol does not use any classical communication,
with its optimal performance.

{\it Symmetry and irreducible decomposition --}
In this letter, we focus on the two kinds of symmetries.
First, our input,
$n$ copies $|\phi\rangle^{\otimes n}$ of $|\phi\rangle$,
is invariant by the reordering of copies, or 
the action of the permutation $\sigma$ in the set $\{1,\ldots n\}$ 
such that,
\begin{align}
           \bigotimes_{i=1}^n|h_{i,x}\rangle
       \mapsto 
           \bigotimes_{i=1}^n|h_{\sigma^{-1}(i),x}\rangle,
\Label{sym}
\end{align}
where $|h_{i,x}\rangle \in {\cal H}_x\;\;(x=A,B)$.
Second, 
a natural protocol will be symmetric with respect to
the action of local unitary transform 
$U^{\otimes n}{\otimes}V^{\otimes n}$ $(U, V\in \SU(d))$,
because 
the Schmidt basis of $|\phi\rangle$ is unknown,
and universal concentration protocol
must be prepared for all the possibility.

Action of these groups
occurs a decomposition of 
the tensored space ${\cal H}_x^{\otimes n}(x=A, B)$ 
~\cite{Weyl,GW,Iwa},
\begin{align}
{\cal H}_x^{\otimes n}=\bigoplus_{{\bf n}}{\cal W}_{{\bf n},x} ,\;
{\cal W}_{{\bf n},x}:= {\cal U}_{{\bf n},x} \otimes {\cal V}_{{\bf n},x}\; (x=A, B) 
\Label{h4}
\end{align}
where 
${\cal U}_{{\bf n},x}$ and 
${\cal V}_{{\bf n},x}$ 
is an irreducible space of 
the tensor representation of $\SU(d)$, 
and the representation~(\ref{sym}) of 
the group of permutations respectively,
and 
\begin{align*}
{\bf n} = (n_1, \ldots, n_d) ,\quad
\sum_{i=1}^d n_i= n,\; n_{i} \ge n_{i+1}\ge 0,
\end{align*}
is called {\it Young index}, which 
${\cal U}_{{\bf n},x}$ and 
${\cal V}_{{\bf n},x}$
uniquely correspond to.
In the case of spin-$\frac{1}{2}$-system, ${\cal W}_{{\bf n}, x}$ is
an eigenspace of the total spin operator, with corresponding
eigenvalue $\frac{(n_1-n_2)(n_1-n_2+2)}{4}$. 

Due to the invariance by the permutation (\ref{sym}),
and with the help of lemmas~\ref{lem:decohere}-\ref{lem:shur}
(see appendix for more detail),
any $n$-tensored state
$|\phi\rangle^{\otimes n}$
is written as,
\begin{align}  
|\phi\rangle^{\otimes n}
=\sum_{\bf n} |\phi_{{\bf n}}\rangle \otimes |{\cal V}_{\bf n}\rangle.
\Label{decom2}
\end{align}
Here, $|\phi_{{\bf n}}\rangle$ is a state 
in ${\cal U}_{{\bf n}, A}\otimes {\cal U}_{{\bf n}, B}$,
which is dependent on $|\phi\rangle$, while 
$|{\cal V}_{\bf n}\rangle $ is $\dim {\cal V}_{{\bf n}, A}$-dimensional
maximally entangled state in  
${\cal V}_{{\bf n}, A}\otimes {\cal V}_{{\bf n}, B}$,
which is independent of $|\phi\rangle$.
Our goal is to pull out  maximally entangle states 
$|{\cal V}_{\bf n}\rangle$, which are 
embedded in any $n$ copies of entangled pure states.

\vspace*{1mm}
{\it The protocol $\{C_*^n\}$--} 
Now, we present the protocol $\{C_*^n\}$, in which 
$C_*^n$ allows $n$ copies of an arbitrary pure entangled state
as its input,
outputting a perfect $2^{nH({\bf p}_{\phi})}$-dimensional
maximally entangled state with almost always in case $n\gg 1$.
The protocol consists of three operations $O1$, $O2$, and $O3$.
( The operation $O1$ is needed only to simplify the mathematical analysis,
  and the protocol performance is not effected by 
 this operation.)

First, both parties independently and randomly choose a unitary transform
$U, V$  according to the uniform distribution 
(Haar measure in $\SU(d)$),  
apply $U^{\otimes n}$, $V^{\otimes n}$  to their particles,
and erase the information about $U, V$ (operation $O1$).

By virtue of lemmas~\ref{lem:decohere}-\ref{l5},  
the average state  is written as,
(see the end of the letter for more explanation),
\begin{align}
&{\rm E}_{U, V} 
(U\otimes V 
|\phi\rangle\langle\phi|
 U^* \otimes V^*)^{\otimes n}
= \sum_{\bf n} 
a_{\bf n}^{\phi}
\sigma_{\bf n}^{\phi},
\Label{5-5-1} \\
& a_{\bf n}^{\phi}:=
\Tr\left\{
{\cal W}_{{\bf n},A}\left(\Tr_B |\phi \rangle \langle \phi|\right)^{\otimes n}
\right\}, \nonumber \\
& \sigma_{\bf n}^{\phi}
=
\frac{1}
{\dim {\cal U}_{{\bf n},A}\otimes {\cal U}_{{\bf n},B}} 
{\cal U}_{{\bf n}, A}\otimes {\cal U}_{{\bf n}, B} 
\otimes|{\cal V}_{\bf n}\rangle\langle {\cal V}_{\bf n}|,\nonumber
\end{align}
where the projection onto a Hilbert space  
${\cal X}$ is denoted  also by ${\cal X}$, and
${\rm E}_{U, V} $ denotes the expectation concerning the
uniform distribution of $U$ and $V$.

The state~(\ref{5-5-1}) is
a probabilistic mixture of the states
$\sigma_{\bf n}^{\phi}$
and each of these states is supported on 
${\cal W}_{{\bf n}_A,A} \otimes{\cal W}_{{\bf n}_B,B}$,
which is orthogonal with each other.
Hence, Alice's measurement 
$\{ {\cal W}_{{\bf n}_A,A}\}_{{\bf n}_A}$ and 
Bob's measurement $\{ {\cal W}_{{\bf n}_B,B}\}_{{\bf n}_B}$
yield the same measurement result  
${\bf n}_A={\bf n}_B={\bf n}$, and the state is changed to
$\sigma_{\bf n}^{\phi}$ (operation $O2$).

Tracing out unnecessary part
${\cal U}_{{\bf n},A}
\otimes 
{\cal U}_{{\bf n},B}$ 
of the state~$\sigma_{\bf n}^{\phi}$, we finally obtain 
$|{\cal V}_{\bf n}\rangle$, 
which is a maximal entangled state with dimension
$\dim{\cal V}_{{\bf n}, A}=\dim{\cal V}_{{\bf n}, B}$,
with  the probability 
$a_{\bf n}^{\phi}$.
(operation $O3$).
Our protocol $C_*^n$ obviously satisfies the 
distortion-free property.

{\it Performance of $\{C_*^n\}$ -- }
A universal entanglement concentration $\{C^n\}$ is a sequence of 
LOCC measurements $C^n$, which allows $n$ copies 
$|\phi\rangle^{\otimes n}$  of unknown state as its input.
Each $C^n$
outputs 
$\rho^{\phi}_{C^n}(L)$, 
which is approximate to
$L$-dimensional
maximally entangled state $\|L\rangle$,
together with $L$ as a classical information, 
with the corresponding probability $Q^{\phi}_{C^n}(L)$.

To fit this formalism, our protocol $\{C_*^n\}$ needs one more
operation which changes $|{\cal V}_{\bf n}\rangle$ to
$\|\dim{\cal V}_{\bf n}\rangle$. 
After this modification, 
$\rho^{\phi}_{C_*^n}(L)=\|L\rangle\langle L\|$ and
$Q^{\phi}_{C^n}(L)=a_{\bf n}^{\phi}$,
if $L=\dim{\cal V}_{\bf n}$ for some ${\bf n}$.
Otherwise, $Q^{\phi}_{C^n}(L)=0$.

The distortion $\epsilon^{\phi}_{C^n}$ of the protocol $C^n$ is defined as
the maximum of the square of the Bures' distance between 
the output  $\rho^{\phi}_{C^n}(L)$ and 
the target $\|L\rangle$,
\begin{align}
 \epsilon^{\phi}_{C^n}
:=1- \min_{L}\langle L\|
\rho^{\phi}_{C^n}(L)
\|L\rangle.
\end{align}
In a universal approximate (distortion -free) concentration,
$\lim_{n\to\infty}\epsilon^{\phi}_{C^n}=0$ ($\epsilon^{\phi}_{C^n}=0$).

We denote $\sum_{L< S} Q^{\phi}_{C^n}(L)$, or 
the probability 
that the protocol fails to produce
maximally entangled state more than $L$,
by $P^{\phi}_{C^n, S}$.
For main difficulty of universal concentration comes from
lack of information about Schmidt basis,
we consider
the failure probability  of the worst case with respect to Schmidt basis,
\begin{align}
\max_{U, V}
  P^{U\otimes V\phi}_{C^n, S},
\Label{k2}
\end{align}
where $U$ and $V$ run all over unitary matrices.
For the quantity~(\ref{k2}) decreases
exponentially as $n\to\infty$ (for good protocols), 
the asymptotic behavior of 
the protocol performance is nicely characterized by 
the exponent
\begin{align}
\varlimsup
\frac{-1}{n}\log\max_{U,V}
P^{U \otimes V \phi}_{C^n,2^{nR}}\quad.
\Label{exponent}
\end{align}
Using eqs.~(\ref{l3}), 
the  exponent for $\{C_*^n\}$ 
is calculated as,
\begin{align}
\lim_{n\to\infty}
\frac{-1}{n}\log \max_{U,V}
P^{U \otimes V \phi}_{C_*^n,2^{nR}}
=
\min_{H( {\bf q}) \le R}
D({\bf q}\| {\bf p}_{\phi})
\Label{fexp1-1},
\end{align}
where $D({\bf q}\| {\bf p}_{\phi}):=\sum_i q_i\log\frac{q_i}{p_i}$
is  classical relative entropy.

Eq.~(\ref{fexp1-1}) implies that our protocol achieves entropy rate: 
if $R$ is strictly smaller than $H({\bf p}_{\phi})$, 
the failure probability goes to zero,
for the r.h.s. of eq.~(\ref{fexp1-1}) is positive;
\begin{thm}
 Our protocol produces
the $2^{nH({\bf p}_{\phi})}$-dimensional 
maximally entangled state with probability one, asymptotically,
and is distortion-free.
\Label{rate}
\end{thm}

{\it Optimality of $\{C_*^n\}$--}
Though there 
are many kinds of 
universal approximate/distortion-free concentration protocols,
our protocol $\{C_*^n\}$ is optimal;
\begin{thm}
Our protocol $C_*^n$ achieves the optimal value of eq.~(\ref{k2}) for
all universal distortion-free concentrations for all finite $n$.
\Label{opt1}
\end{thm}
\begin{thm}
Our protocol $\{C_*^n\}$ attains the optimal value of 
eq.~(\ref{exponent}) for
      all universal approximate concentrations.     
\Label{optapprox}
\end{thm}

Note that the optimality result of theorem~\ref{opt1} is non-asymptotic
and holds for any finite number $n$, while 
theorem~\ref{optapprox} is an  asymptotic result.
Note also that 
the l.~h.~s. of eq.~(\ref{fexp1-1}) is the same with 
that of the  protocol in
\cite{Ben}, 
which  
{\it does} 
use the knowledge of the Schmidt basis of the entangled pairs.

\begin{lem}
Without loss of generality, we can restrict ourselves to
the improvement of  $\{C_*^n\}$ by some post processing,
or the protocols of  the form  $\{C^n_*+B^n\}$,
where $B^n$ is an LOCC operation.
\Label{optimal}
\end{lem}

Here, the operation $O+O'$ means  doing  the operation $O$  and 
the operation $O'$ in succession.

\begin{pf}
It suffices to prove that 
all the action of our protocol do not disturb 
procedures which might follow:
given an arbitrary 
universal distortion-free(approximate) concentration $\{B^n\}$,
eq.~(\ref{k2}) of $C_*^n+B^n$ is not larger than that of $B^n$.

The operation $O1$
will not increase the failure (\ref{k2}), because
the protocol must be prepared for all kinds of Schmidt bases.
Indeed, whatever the unitaries $U', V'$ are applied, 
\begin{align*}
\max_{U, V}
  {\rm E}_{U', V'}
   P^{U U'\otimes V V'\phi}_{B^n, S} 
=  {\rm E}_{U', V'}
   P^{U'\otimes V'\phi}_{B^n, S} 
\leq
\max_{U, V}  P^{U \otimes V \phi}_{B^n, S},
\end{align*}
and hence random  application of local unitaries does not
increase the failure. 
In the same way, it is also shown that  
$O1+B^n$ is also distortion-free (approximate)
if $B^n$ is distortion-free (approximate).

The second stage, or the local measurements 
$\{{\cal W}_{{\bf n}, A}\}$ and $\{{\cal W}_{{\bf n}, B}\}$,
will not cause distortion, 
because the state is already block diagonal in 
subspaces $\{{\cal W}_{{\bf n}, A}\otimes{\cal W}_{{\bf n}', B}\}$.
Tracing out  
${\cal U}_{{\bf n},A}\otimes {\cal U}_{{\bf n},B}$, 
which is separable,
will not affect the further operations, 
because these states can be 
reproduced by LOCC whenever they are needed.
Therefore, $C_*^n+B^n$ is also distortion-free(approximate),
and its performance is not worse than $B^n$.
\end{pf}

For  the performance of protocols of the type
stated in lemma~\ref{optimal}
is symmetric,
$
\max_{U,V}  
P^{U \otimes V \phi}_{C^n, S}
=
P^{\phi}_{C^n, S}$,
there is no need to take maximum
of the failure probability in the following.

To improve $C_*^n$ by post processing, 
it is needed to transform smaller dimensional
maximally entangled state $\|L\rangle$ to larger one $\|M\rangle$,
$(M>L)$, exactly or approximately.
Here, the monotonicity of Schmidt rank by LOCC leads to 
a significant fact:
one of the best approximate states to $\|M\rangle$ 
which can be generated
from $\|L\rangle$ is $\|L\rangle$ itself.
(Note that we can concentrate on transform to 
 pure state,  
for any transition to mixed state can be considered to be
probability mixture of transitions to  pure states,
as in \cite{VidalJonathanNielsen}.)
Hence, we have; 
\begin{lem}
An optimal post processing is as follows.
Do not change the output  state, and change 
the 'label':  claim $\|L\rangle$
to be an approximate state to a larger maximally entangled
state $\|M\rangle$ with some corresponding probability.
\Label{lem:k1}
\end{lem}

This lemma directly yields theorem~\ref{opt1}.
For the proof of  theorem~\ref{optapprox}, 
the problem is related to
the statistical estimation of $H({\bf p}_{\phi})$,
and the results in statistics is made use of.
Roughly speaking,
the asymptotic performance of $\{C^n\}$ is bounded by
that of the optimal estimate of $H({\bf p}_{\phi})$,
because entanglement concentrated by $C^n$ gives 
natural estimate of $H({\bf p}_{\phi})$
when $n\gg 1$.

More precisely, we consider 
the estimation of $H({\bf p}_{\phi})$ 
from the classical information $L$,
assuming that $L$ obeys probability distribution
$Q^{\phi}_{C_*^n}(L)$, with 
$|\phi\rangle$ being unknown:
by a post processing as in lemma~\ref{lem:k1},
classical output $L$ of $C_*^n$ is  changed 
to $M$ with some corresponding probability,
by which the estimate of  $H({\bf p}_{\phi})$ is calculated as 
$\frac{1}{n}\log M$.

\vspace*{1mm}
\noindent
{\bf Proof of theorem~\ref{optapprox}}\quad
{\it Strong converse theorem} of entanglement concentration assures
the probability of 'too much success',
or achieving the rate more than the entropy of entanglement,
tends to zero~\cite{LP, Morikoshi}.
Therefore, if $\{C^n\}$ is a universal approximate
entanglement concentration, which means
$\{C^n\}$ achieves the entropy rate,
the concentrated entanglement 'converges to' 
(,thus giving nice estimate of), the entropy of entanglement.
More rigorously, choosing $R$, $|\phi\rangle$, $|\psi\rangle$ so that
$H({\bf p}_{\phi})<R<H({\bf p}_{\psi})$,
we have,
\begin{equation}
P^{\phi}_{C^n, 2^{nR}}(:=p_n)\rightarrow 0,
\quad
P^{\psi}_{C^n, 2^{nR}}(:=q_n) \rightarrow 1.
\Label{converse}
\end{equation}

Therefore,  as is proven later, 
we have Bahadur-type inequality~\cite{Bahadur},
\begin{align}
\mbox{eq.~(\ref{exponent})}\leq 
\varlimsup\frac{1}{n}{\rm D}(Q^{\psi}_{C_*^n}\|Q^{\phi}_{C_*^n} ),
\Label{bahadur2}
\end{align}
whose r.~h.~s. is evaluated
by use of eqs.~(\ref{l3}) as
to be ${\rm D}({\bf p}_{\psi}\|{\bf p}_{\phi})$.
Therefore, choosing 
$|\psi\rangle$ so that $H({\bf p}_{\psi})$
is infinitely close to $R$, 
it is proved that
eq.~(\ref{exponent}) is smaller than or equal to
the r.~h.~s. of eq.~(\ref{fexp1-1}),
and we have the theorem.

Eq.~(\ref{bahadur2}) is proven as follows~\cite{Bahadur}.
Monotonicity of relative entropy implies,
\begin{align*}
&{\rm D}(Q^{\psi}_{C_*^n}\|Q^{\phi}_{C_*^n} )
\geq
{\rm D}(Q^{\psi}_{C^n}\|Q^{\phi}_{C^n} )\\
&\geq
q_n\log \frac{q_n}{p_n} +
(1-q_n)\log \frac{1-q_n}{1-p_n},
\end{align*}
which yields,
\begin{align*}
&\frac{-1}{n}\log {p_n}
=\frac{-1}{n}\log
P^{\phi}_{C^n, 2^{nR}}\nonumber\\
&\leq 
\frac{1}{n q_n}
 \left(
{\rm D}(Q^{\psi}_{C_*^n}\|Q^{\phi}_{C_*^n} ) 
 +h(q_n) +(1-q_n)\log (1-p_n)\right),
\end{align*}
with $h(x):=-x\log x-(1-x)\log(1-x)$.
With the help of eqs.~(\ref{converse}),
letting $n\to\infty$ of the both sides of this inequality, 
eq.~(\ref{bahadur2}) is obtained. 
\endproof

{\it Conclusions and discussions --}
We have proposed a new kind of entanglement concentration, 
a universal distortion-free concentration, 
and constructed an example, $\{C_*^n\}$, 
which turned out to be optimal not only for all
universal distortion-free concentrations, but also
for all universal approximate concentrations.
Remarkably, our protocol does not use any classical
communication.

It is also notable that our protocol gives best estimate of 
$H({\bf p}_{\phi})$, not only for all LOCC measurements but
also for all (possibly global) measurements,
in case that
the Schmidt basis is unknown:
lemma~\ref{optimal} is true in this case too, and 
information about the input state reflects only in 
$Q_{C_*^n}^{\phi}$, for the output state $|{\cal V}\rangle$
is independent of the input.

In the proof of lemmas~\ref{optimal}-\ref{lem:k1},
which are keys to the proof of optimality,
invariance of the performance measure by 
local unitary operations is the only assumption which is made use of.
Therefore, our protocol is optimal 
also in terms of other invariant performances measures, such as 
(\ref{k2})+$\max_{U,V}\epsilon^{U\otimes V\phi}_{C^n}$.


\vspace*{2mm}
{\it Appendix  -- Group representation theory  --}
\begin{lem}
\Label{lem:decohere}
Let  $U_g $  and $U'_g$ be  an irreducible representation
of $G $ on the finite-dimensional space ${\cal H}$ and ${\cal H}'$, respectively.  We further assume that $U_g$ and $U'_g$ are
not equivalent.
       If a linear operator $A$ in ${\cal H}\oplus{\cal H}'$  
       is invariant 
       by the transform 
       $A\to U_g\oplus U'_g  A U_g^*\oplus U_g^{'*}$ for any $g$,
       ${\cal H}A{\cal H'}=0$.
~\cite{GW}
\end{lem}
\begin{lem}
\Label{lem:shur}
(Shur's lemma~\cite{GW}) 
Let $U_g$ be as defined in lemma~\ref{lem:decohere}.
If a linear map $A$ in ${\cal H}$ is invariant 
       by the transform 
       $A\to U_g  A U_g^*$ for any $g$,
       $A=c \Id$.
\end{lem}
\begin{lem}\Label{l5}
If  the representation $U_g (U'_h)$
of the group $G (H)$ on the finite-dimensional space ${\cal H} ({\cal H}')$
is irreducible,
the representation $U_g\times U'_h$
of the group $G\times H$ in the space ${\cal H}\otimes {\cal H}'$
is also irreducible.
\end{lem}
\begin{pf}
Let ${\cal K}$ be an irreducible subspace of ${\cal H}\otimes{\cal H}'$.
Denoting Haar measure in $G$ and $H$ by  $\mu({\rm d}g)$ and $ \nu({\rm d}h)$ respectively,
we obtain 
\begin{align}
&\int U_g\otimes U'_h 
X
U_g^*\otimes U_h^{'*}
\mu({\rm d}g) \nu({\rm d}h)
= const.\times \Id_{{\cal H}\otimes{\cal H}'}.
\Label{hoge}
\end{align}
In case of $X= |\phi_1 \otimes \phi_1' \rangle
\langle \phi_2 \otimes \phi_2'|$, 
follows from the invariance of the r.~h.~s. by both $U_g\cdot U_g^*$
and $U'_h\cdot U_h^{'*}$.
Since any matrix on ${\cal H}\otimes{\cal H}'$ is written
as a sum of the above type of matrices,
eq.~(\ref{hoge}) holds for any matrix.
When $X= \Id_{\cal K}$, the r.~h.~s. of eq.~(\ref{hoge})
equals to $\mu(G)\nu(H) \Id_{\cal K}$.
Since $\mu(G)\nu(H) \neq 0$, we obtain 
${\cal K}= {\cal H}\otimes{\cal H}'$, i.e., 
${\cal H}\otimes{\cal H}'$ is irreducible.
\end{pf}

\noindent
{\bf Group representation type theory}\cite{KW,HMe,Ha}\quad
The following inequality and  equation
in the Appendix of \cite{HMe} and \cite{Ha}, are useful
in the calculation of exponents.
\begin{equation}
\begin{split}
\left|\frac{1}{n}\log \dim {\cal V}_{{\bf n}} - H\left( \frac{{\bf n}}{n}
\right)\right|
\le \frac{d^2 +2d}{2n}\log (n+d), \\
\lim_{n \to \infty}
\frac{-1}{n} \log \sum_{\frac{{\bf n}}{n} \in {\cal R}}
\Tr {\cal W}_{{\bf n}} \rho^{\otimes n}
=
\max_{ {\bf q} \in {\cal R}}
D( {\bf q}\| {\bf p}), 
\Label{l3}
\end{split}
\end{equation}
where 
${\bf p}=(p_1, p_2, \cdots, p_d)$ are the eigenvalues of $\rho$, 
and 
${\cal R}$ is an arbitrary closed subset
of $\{ {\bf q}| q_1 \ge q_2 \ge \ldots \ge q_d\ge 0, \sum_{i=1}^d q_i =1 \}$.
\vspace*{1mm}
\noindent

\noindent
{\bf Sketch of proof of eq.~(\ref{decom2})}\quad
Establish a correspondence 
between $|\phi\rangle$ and  the operator 
$\Phi:= \sum_i \sqrt{p_{i,\phi}}|e_{i,A}\rangle \langle e_{i,A}|$ 
in ${\cal H}_A$. 
Letting $\Phi_{{\bf n}}$ be a linear transform 
in ${\cal U}_{{\bf n}}$, 
$\Phi^{\otimes n}$ is written as 
~\cite{GW, Iwa}, 
\begin{align}  
\Phi^{\otimes n}=\bigoplus_{\bf n} 
\Phi_{{\bf n}}\otimes \Id_{{\cal V}_{{\bf n}}}
\Label{decom}
\end{align}
which is equivalent to eq.~(\ref{decom2}),
corresponding $|\phi_{{\bf n}}\rangle$ 
($\|{\cal V}_{\bf n}\rangle\rangle$) to
$\Phi_{\bf n}$ ($\Id_{{\cal V}_{{\bf n}}}$), respectively.

Lemma~\ref{lem:decohere} assures that 
the 'off-diagonal part'  ${\cal W}_{{\bf n}, A}\Phi^{\otimes n}{\cal W}_{{\bf n}', A}$ 
is zero.
The factor $\Id_{{\cal V}_{{\bf n}}}$
in eq.~(\ref{decom})
is due to the invariance of an input state $|\phi\rangle^{\otimes n}$
by the action (\ref{sym}) of permutation 
(Shur's lemma, lemma~\ref{lem:shur}).
Two spaces ${\cal V}_{{\bf n}, A}\otimes {\cal V}_{{\bf n}, B}$
and ${\cal U}_{{\bf n}, A}\otimes {\cal U}_{{\bf n}, B}$
are disentangled, because 
${\cal V}_{{\bf n}, A}$ and ${\cal V}_{{\bf n}, B}$
are maximally entangled.
\endproof
\vspace*{1mm}
\noindent

\vspace*{1mm}
\noindent
{\bf Sketch of proof of eq.~(\ref{5-5-1})}\quad
Lemma~\ref{l5} assures that 
${\cal U}_{{\bf n}, A}\otimes{\cal U}_{{\bf n}, B}$
is an irreducible space of the tensored representation
$U^{\otimes n}\otimes V^{\otimes n}$ of $\SU(d)\times\SU(d)$.
Lemmas~\ref{lem:decohere}-\ref{lem:shur},
letting $G$ be $\SU(d)\times\SU(d)$,
lead to eq.~(\ref{5-5-1}).
\endproof

\end{document}